# Exotic Hadrons

*Francesco Renga[1]*

(1) "Sapienza" Università di Roma and INFN Sezione di Roma, P.le Aldo Moro 2, 00158 Roma (Italy), francesco.renga@roma1.infn.it

**Abstract**

I review here the most recent results about the observation and the study of hadronic bound states that do not fit well in the standard quarkonium picture. Several new states have been observed in the last few years, at *B*-, τ-Factories and hadron colliders. For most of them, quantum number determinations are available and allow to develop the basis of a new spectroscopy based on exotic compounds like tetraquarks or meson molecules. Nonetheless, there is still a lot of work to do to complete the picture.

**Introduction**

The quarkonium is the bound state of a quark and an anti-quark. The most popular example is the charmonium (c anti-c). Apart from its flavour content and its mass, it is usually classified according to a principal quantum number ($n$), the orbital ($L$), spin ($S$) and total angular momentum ($J$), its parity ($P$) and its charge conjugation parity ($C$). The latter two are related the the former ones by $P = (-)^{L+1}$ and $C = (-)^{L+S}$. Finally, the combination $J^{PC}$ is often used for classification purposes.

The quarkonium can be produced through several mechanisms. For instance, at an $e^+e^-$ collider, it can arise from $e^+e^-$ annihilation through the exchange of an s-channel photon, $e^+e^- \rightarrow \gamma^* \rightarrow$ (q anti-q), possibly with the emission of an initial state radiation (ISR) photon to get the correct center-of-mass energy. At the same machines, it can be also produced from two-photon fusion $e^+e^- \rightarrow e^+e^- \gamma^*\gamma^* \rightarrow e^+e^-$ (q anti-q), double quarkonium production $e^+e^- \rightarrow$ (q anti-q)(q anti-q) or in the decay of other particles, like a B meson.

Finally, the dominant decay mechanism of the quarkonium depends primarily on its mass: below the open flavour threshold (the D anti-D threshold for the charmonium) the decay occurs through electromagnetic or OZI suppressed decays, resulting in a quite stable state with a narrow width. Conversely, above the threshold, the OZI allowed processes dominates, generating wider resonances.

From the discussion above, it emerges clearly that the standard quarkonium states satisfy a series of requirements, in terms of quantum numbers and decay properties, and we can identify as *exotica* the states that violates these simple rules.

In the last few years, *B*-, τ-Factories and hadron colliders allowed to produce and reconstruct several resonances that in fact do not fit well in the standard picture. Their decay products suggest a charmonium-like quark content, but their mass are far from the predictions of the charmonium models. Moreover, although they lie above the D anti-D threshold, their width are often quite small and the D anti-D decay modes seems to be suppressed for some of them. The number of these states is continuously growing. For a recent review, see [1].

A few kinds of bound states, predicted by QCD, could possibly explain these behaviours, but their existence has not been yet proved definitively. They include tetraquarks, meson molecules and hybrids. Tetraquarks (see for instance [2]) are bound states of a diquark (qq) and an anti-diquark (anti-q anti-q). They can have small widths even above the open flavour threshold and a large number of states are expected, typically organised in multiplets with peculiar mass patterns. An example is shown in Fig. 1. Moreover, charged states are predicted.

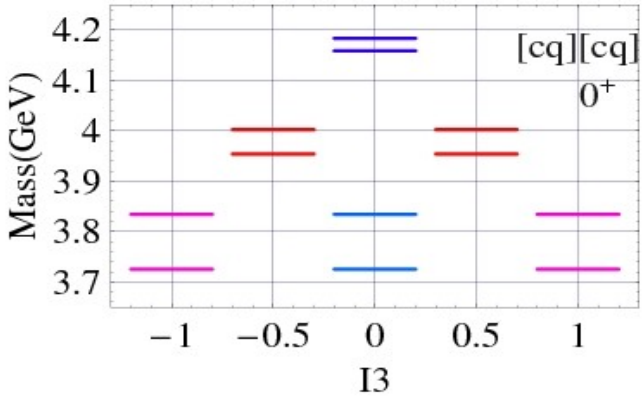

***Fig. 1:*** *Predicted mass pattern for the 0+, L = 0 (cq)(anti-c anti-q) tetraquarks [1].*

Molecular D anti-D states are also predicted by many authors (see for instance [3]). Also in this case small widths can be explained, but a much smaller number of sates is expected. Finally, hybrid states composed by a q anti-q pair and a gluon are also predicted (see [4]). The corresponding resonances would decay mainly to D anti-D pairs and could have quantum numbers forbidden for standard quarkonia, like $J^{PC} = 1^{-+}$.

From the experimental point of view, the search for new multiplets, the search for specific decay channels of the new states and the determination of the quantum numbers are needed in order to get a clue about the underlying structures.

## The X(3872)

The first exotic candidate, the X(3872), has been discovered in 2003 by the Belle experiment, at the KEKB B-Factory, in the decay chain B → X K, X → J/ψ π⁺ π⁻ [5], as shown in Fig. 2. This state has been soon confirmed by the BaBar experiment at the PEP-II B-Factory and by CDF and D0 at the Tevatron p anti-p collider. It has been also observed in a few different decay modes like D* anti-D and J/ψ ω, and an angular analysis of the CDF data excluded all quantum number assignments other than $1^{++}$ and $2^{-+}$.

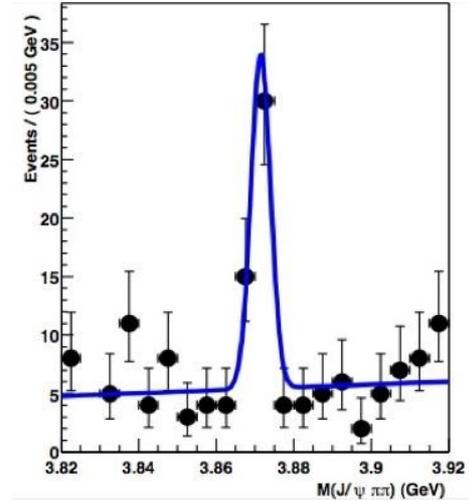

***Fig. 2:*** *The J/ψ π⁺ π⁻ mass spectrum with the X(3872) peak from [5].*

There are several arguments against a regular charmonium interpretation of the X(3872). At first, the width measured by BaBar [6] is $(3.0^{+1.9}_{-1.4} \pm 0.9)$ MeV, much smaller than the one expected for a regular charmonium above the open charm threshold. Second, the Belle's analysis prefers $J^{PC} = 1^{++}$, but the X mass, measured to be $(3872.2 \pm 0.4)$ MeV, is far from the charmonium prediction for this quantum number combination. Moreover, as shown in [1] and in Fig. 3, the B → X K branching ratio (~ $10^{-4}$) is unusually low for a charmonium state. Finally, there is some tension between the mass measurements in the J/ψ π⁺ π⁻ and in the D* anti-D modes, that could indicate the presence of two different states belonging to an exotic (e.g. tetraquark) multiplet. The tension is currently at the level of 3.5σ, but the mass measurement in the D* anti-D mode could be biased by the nearby D* anti-D threshold, so that a final conclusion cannot be drawn.

There are also a few arguments against a molecular interpretation of the X(3872). Among them, the partial width of X → D* anti-D is of the order of 1 MeV, much larger that the one naively expected for a molecular state (~ Γ(D*) ~ 70 keV).

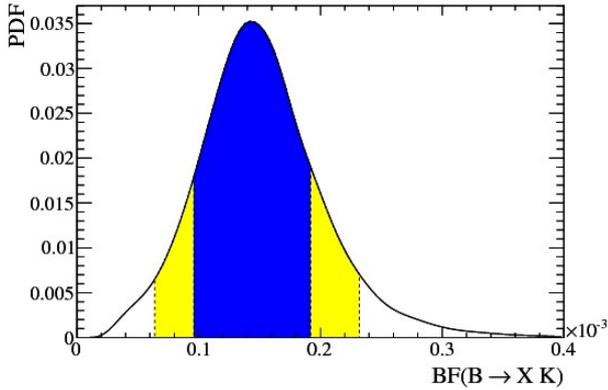

***Fig. 3:*** *The B → X K Bayesian posterior probability obtained in [1] from a combination of all the available X(3872) measurements.*

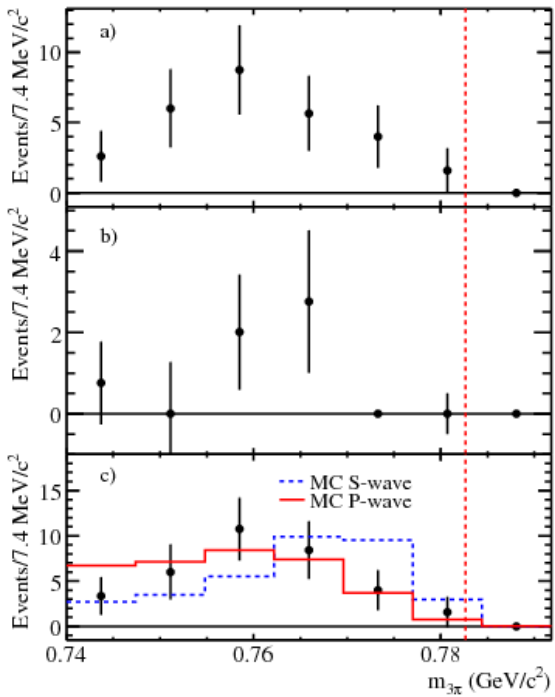

***Fig. 4:*** *The $\pi^+\pi^-\pi^0$ mass distributions for (a) $B^+ \to X K^+$, (b) $B^0 \to X K^0_S$ and (c) the combined distribution, for events with 3.8625 < m(J/ψ ω) < 3.8825 GeV, from [7].*

On the other hand, a recent analysis of the J/ψ ω ($\pi^+\pi^-\pi^0$) mode by BaBar [7] studied the invariant mass distribution of the three pions, showed in Fig. 4, and found a better data description if the X is assumed to have L = 1 (P-wave), so that the $2^{-+}$ assignment is favoured. It would make possible an identification of the X(3872) with the regular charmonium $\eta_{c2}(1D)$.

## The 3940 Family

As mentioned in the Introduction, spotting some multiplet of exotic states could give some important information about the interpretation of these new resonances. An interesting multiplet candidate is the so-called 3940 family. It is composed by four states, with masses around 3940 MeV, all discovered by the Belle collaboration. The first of such states, the Y(3940), has been discovered in the decay chain B → Y K, Y → J/ψ ω [8]. The Z(3930) → D anti-D, and the Y(3915) → J/ψ ω resonances have been then discovered with a two-photon production mechanism [9,10]. Another state, the X(3940) → D anti-D*, has been observed in a double charmonium production, is association with the J/ψ [11]. The preferred quantum number assignments for the Y(3940) and the Y(3915) is $1^{++}$. The mass measurements for these resonances still have large uncertainties, and the identification of them as a unique state is still possible. This would be the first case of an exotic state produced by two different mechanisms. Concerning the X(3940), only the C = + assignment is permitted by its production mechanism, and J = 0 is preferred because only spin 0 states have been observed so far in double charmonium production. Finally, a recent analysis of the angular distributions in Z → D anti-D by BaBar [12] suggests a spin 2 assignment, in agreement with the standard charmonium $\chi_{c2}(2P)$.

## The 4140 Family

Another interesting family of states lies around 4140 MeV, and is composed by the X(4160), found by Belle in a double charm production, in association with the J/ψ [13], and the Y(4140) produced at CDF in B → Y K [14]. The decay modes are X → D* anti-D* and Y → J/ψ φ. The X(4160), in analogy with the X(3940), has positive charge conjugation, and J = 0 is favoured by the production mechanism, so that a possible regular charmonium assignment is the $0^{-+}$ state $\eta_c(3S)$. Conversely, the Y(4140) admits several $J^{PC}$ assignments. Among them, the $1^{-+}$ is of particular interest, because it is forbidden for a regular charmonium and would allow to identify the Y(4140) as a hybrid state, whose mass is predicted to be not so far from 4140 MeV. Unfortunately, an enhanced production in the two-photon fusion mechanism would be expected in this case, while a specific search performed by Belle did not provide any evidence for it [15]. Instead, a new state, named X(4350), was found in this analysis. Finally, a new analysis of the CDF data suggests the presence of a second peak at 4274 MeV in the J/ψ φ spectrum.

## The $1^{--}$ Family

A set of resonances have been discovered by BaBar and Belle through ISR production, allowing the immediate assignment $J^{PC} = 1^{--}$ (the same of the s-channel photon that mediates the production). The first of them, the Y(4260), has been observed by Belle in Y → J/ψ $\pi^+ \pi^-$ [16]. Another one has been discovered by Belle in the same final state, but with a mass of 4660 MeV [17], and named Y(4660). Finally, BaBar observed the Y(4320) decaying to ψ(2S) $\pi^+ \pi^-$ [18]. These states have been deeply studied, in particular because they are good candidates for a tetraquark interpretation, and this hypothesis can be tested by studying their decay modes. For a regular $1^{--}$ charmonium, the favoured decay channels should be D(*) anti-D(*). Conversely, tetraquarks are expected to have large branching ratios to baryonic modes like $\Lambda_c \Lambda_c$. So far, only upper limits are available for the Y → D(*) anti-D(*) modes, the most stringent being BR(Y → D anti-D)/BR(Y → J/ψ $\pi^+ \pi^-$) < 1 at 90% C.L. for the Y(4260). Conversely, as shown in Fig. 5, a large signal for Y(4660) → $\Lambda_c \Lambda_c$ has been already observed [19], making it the best tetraquark candidate available so far.

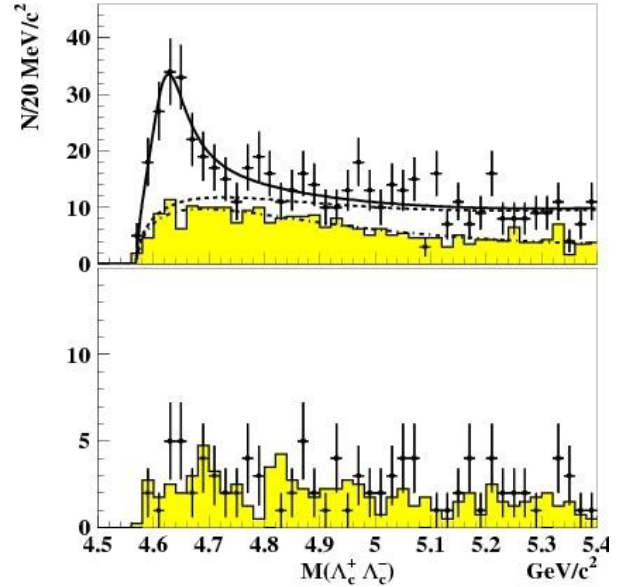

**Fig. 5:** *(Top) the $\Lambda_c \Lambda_c$ spectrum and (bottom) a control sample for checking the background, from [19].*

## Charged States

As already mentioned, the tetraquark model predicts also charged charmonium-like states, that are not predicted by other models and in particular could not be interpreted in any way as regular charmonia. It stimulated the search for charged states, but so far only Belle claimed an observation. In particular, Belle observed a resonance, the $Z^+(4430)$, produced in B → Z K and decaying into ψ(2S) $\pi^+$ [20], as shown in Fig. 6. Moreover, two structures were observed at 4050 and 4250 MeV in the B → Z K, Z → $\chi_{c1}$ π decay chain, and named $Z_1$ and $Z_2$ [21]. There has been anyway some criticism about the Belle analysis that produced the observation of the Z(4430). In fact, Belle initially desisted from a complete treatment of the three body dynamics of the K ψ(2S) $\pi^+$ system, and simply applied an invariant mass cut to remove Kπ resonances.

Conversely. BaBar published an analysis [22] where a deep study of the Kπ background, mainly based on control samples from real data, was performed. Although the BaBar data are statistically consistent with the Belle's one, BaBar did not found any evidence for a resonance at 4430 MeV, but only some hint for a peak at 4476 MeV, 3σ away from the Belle's one and with only 2.9σ of significance, so that the Belle's observation is still unconfirmed. Following the BaBar's analysis, Belle performed a full Dalitz plot analysis of its data [23]. The result of the previous analysis is essentially confirmed, but a much larger error on the mass estimate is obtained, making it consistent with the BaBar peak position.

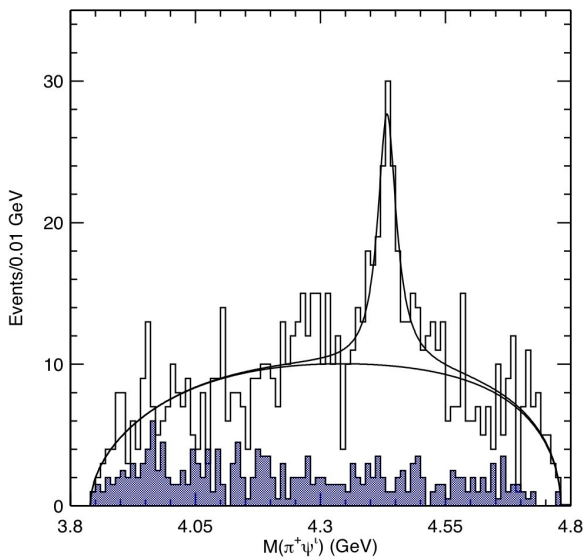

*Fig. 6:* The ψ(2S) π+ spectrum from [20] with the fitted $Z^+(4430)$ peak.

## Exotic Bottomonium

The discovery of several candidates for charmonium-like exotic resonances triggered a search for their bottom companions. The easier way to look for such states at the B-Factories is the search for 1⁻ resonances in $e^+e^-$ annihilation above the B anti-B threshold. A naïve mass scaling, according to the mass difference of the regular 1⁻ charmonia and bottomonia, can be applied to the resonances of the 1⁻ family, and suggests the presence of exotic bottomonia just above the Y(4S) resonance. The interest in this field has been boosted by the Belle's observation of an anomalously large Y(nS) π+π− production around the Y(5S) [24], as shown in Fig. 7. This decay channel is the bottom companion of the J/ψ π+ π− mode, preferred by several exotic charmonia. It suggests the possibility of an exotic state in the same energy region, or the interpretation of the Y(5S) itself as an exotic resonance [25].

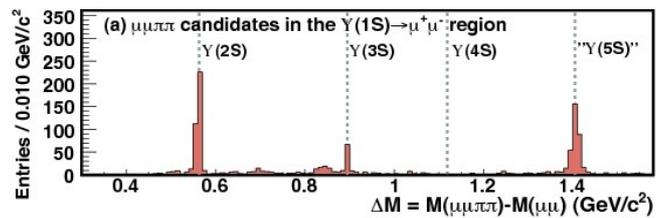

*Fig. 7:* The distribution of ΔM = M(Y(1S) π π) − M(Y(1S)) from [25], showing the large contribution for M(Y(1S) π π) = M(Y(5S)).

Following this result, both BaBar and Belle performed an energy scan in the region of the B(*) anti-B(*) thresholds. The BaBar scan [26], from 10.54 to 11.02 GeV in steps of 5 MeV, revealed several structures in the inclusive $e^+e^- \to$ b anti-b cross section (see Fig. 7), but their interpretation is made difficult by the presence of threshold and final state interaction effects. In fact, most of these structures can be qualitatively explained without including any exotic resonance [27].

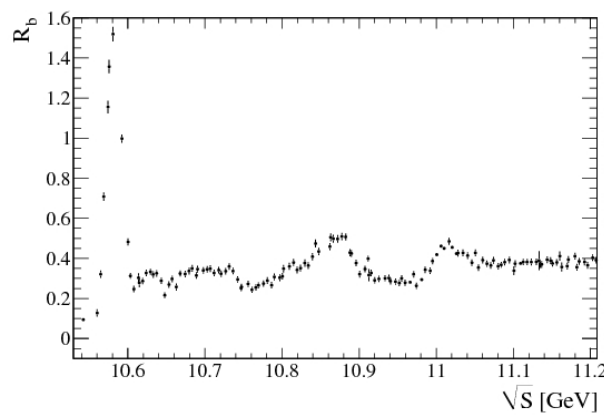

*Fig. 8:* The $e^+e^- \to$ b anti-b cross section normalized to the $e^+e^- \to \mu^+\mu^-$ cross section as obtained in [26].

The scan also allowed an estimate of the Y(5S) and Y(6S) shapes, but the mass and width measurements obtained in this way can be strongly affected by final state interactions that prevent to extract the pole values from a simple fit of the inclusive shape with a superposition of Breit-Wigner resonances and continuum contributions.

Belle concentrated its data taking in few points around the Y(5S), but collected a larger statistics for each scan point, allowing a precise measurement of the exclusive Y(nS) $\pi^+\pi^-$ rates and shapes [28]. The Y(5S) mass and width extracted in this way are 2.2σ and 1.4σ away from the BaBar one. Once again, the interpretation of any discrepancy of this kind is made difficult by the different impact that final state interactions have on the different channels [29, 30]. A more comprehensive theoretical description, including both inclusive and exclusive decay modes, would be needed in order to finally assess the nature of the Y(5S).

## Conclusions and Outlook

I reviewed the most recent results in the search for exotic hadronic states. Several states that do not fit in the standard charmonium picture have been already discovered and also the quantum numbers are known for some of them. Anyway, only a small fraction of the possible final states have been searched for and only a small fraction of the available decay spectra has been fitted to set limits on the branching ratios of the known new resonances. Some of these spectra also suffer from low statistics. In conclusion, there is still a lot of work to do with the present data and there is room for improvements if a new generation of experiments, like a Super B-Factory, will be available. Moreover, the LHC experiments will have the opportunity of confirming the present observations and provide new information about the production of charmonium-like states at hadron collider, which could give a clue for their interpretation. Finally, there is still work to do from the theoretical side in order to clarify the picture and possibly build a new spectroscopy.